\documentstyle[emulateapj,psfig]{article}

\makeatletter

\newenvironment{inlinefigure}{%
\def\@captype{figure}%
\noindent\begin{minipage}{0.999\linewidth}\begin{center}}
{\end{center}\end{minipage}\smallskip}
\makeatother

\begin{document}
\title{The Evolution of the Ultraluminous Infrared Galaxy
Population from Redshift 0 to 1.5\altaffilmark{1,2}}
\author{
L.~L.~Cowie,$\!$\altaffilmark{3}
A.~J.~Barger,$\!$\altaffilmark{4,5,3}
E.~B.~Fomalont,$\!$\altaffilmark{6}
P.~Capak$\!$\altaffilmark{3}
}

\altaffiltext{1}{Based in part on data obtained at the W. M. Keck
Observatory, which is operated as a scientific partnership among the
the California Institute of Technology, the University of
California, and NASA and was made possible by the generous financial
support of the W. M. Keck Foundation.}
\altaffiltext{2}{Based in part on data obtained at the VLA, which is 
a facility of the National Radio Astronomy Observatory (NRAO). 
The NRAO is a facility of the National Science Foundation operated
under cooperative agreement by Associated Universities, Inc.}
\altaffiltext{3}{Institute for Astronomy, University of Hawaii,
2680 Woodlawn Drive, Honolulu, HI 96822.}
\altaffiltext{4}{Department of Astronomy, University of
Wisconsin-Madison, 475 North Charter Street, Madison, WI 53706.}
\altaffiltext{5}{Department of Physics and Astronomy,
University of Hawaii, 2505 Correa Road, Honolulu, HI 96822.}
\altaffiltext{6}{National Radio Astronomy Observatory,
Charlottesville, VA 22903.}

\slugcomment{Accepted by The Astrophysical Journal Letters}

\begin{abstract}
We use redshift observations of two deep 1.4~GHz fields 
to probe the evolution of the bright end of the radio galaxy luminosity 
function to $z=1.5$. We show that the number of galaxies 
with radio power that would correspond locally to an ultraluminous
infrared galaxy (ULIG) evolves rapidly over this redshift 
range. The optical spectra and X-ray luminosities are used to refine this 
result by separating the sources with clear active galactic nucleus (AGN)
signatures from those that may be dominated by star formation. Both
populations show extremely rapid evolution over this redshift range. We
find that the number of sources with ULIG radio power and no clear AGN 
signatures evolves as $(1+z)^{7}$.
\end{abstract}


\section{Introduction}
\label{secintro}

Star-forming ultraluminous infrared galaxies, or ULIGs
($L_{FIR}>10^{12}~L_\odot$ from 
\markcite{sanders96}Sanders \& Mirabel 1996;
here we mean galaxies without clear signatures of active galactic nucleus
[AGN] activity in their spectra, even if there may be some AGN 
contribution to their luminosities), represent the most luminous tip of 
the star-forming galaxy population. Locally, ULIGs are very scarce 
(\markcite{kim98}Kim \& Sanders 1998), but if, as is widely believed, 
submillimeter galaxies are the high redshift analogs, then by 
$z=1$ they are the dominant contributors to the
star formation history (e.g., \markcite{lilly99}Lilly et al.\ 1999;  
\markcite{bcr00}Barger, Cowie, \& Richards 2000;
\markcite{gispert00}Gispert, Lagache, \& Puget 2000; 
\markcite{chapman03}Chapman et al.\ 2003). 
It is extremely hard to trace the evolution of the ULIG population
at low redshifts ($z<1.5$) because of the difficulties of 
mapping large areas with current submillimeter
detectors and of identifying the optical counterparts to low spatial 
resolution submillimeter or far-infrared (FIR) observations.  

Ultradeep decimetric radio surveys can play a major, complementary 
role. The well-known tight correlation between
FIR luminosity and radio power in local star-forming galaxies and 
radio-quiet AGNs (\markcite{condon92}Condon 1992)
means that we can identify ULIGs out to substantial redshifts based 
on their radio power. 
For a 1.4~GHz sensitivity limit of 40~$\mu$Jy,
an $L_{FIR}=10^{12}~L_\odot$ source will be detected to $z\sim 1.5$.
Thus, sources with radio power that would correspond locally to a ULIG 
may be seen directly to large redshifts. Since the ultradeep 1.4~GHz 
samples cover large fields, do not suffer from extinction, 
and provide subarcsecond positional accuracy, it is easy to develop 
large samples with highly complete optical counterpart identifications.
By contrast, optically selected samples may omit dusty sources at 
these redshifts, while X-ray samples will select AGN dominated sources.

Current decimetric surveys do not probe the 
normal star-forming galaxy populations at $z=1$. To the deepest 
1.4~GHz sensitivity limits of $\sim 20$~$\mu$Jy ($5\sigma$),
a source with radio power corresponding 
to $L_{FIR}=10^{10}~L_\odot$ would only be seen to $z\sim 0.3$. 
Thus, attempts to provide maps of the entire star formation
history over the $z=0-1$ range based on decimetric observations
(e.g., \markcite{cram98}Cram 1998; 
\markcite{mobasher99}Mobasher et al.\ 1999;
\markcite{haarsma00}Haarsma et al.\ 2000, hereafter H00) 
have to rely on model luminosity 
functions to extrapolate the measured data and obtain star formation rates.
In these descriptions, much of the star formation lies in
sources that are not directly measured by the observations (e.g., H00).

We use ultradeep 1.4~GHz observations of the
Hubble Deep Field-North (HDF-N; \markcite{richards00}Richards 2000) 
and SSA13 (\markcite{fomalont04}Fomalont et al.\ 2004) fields
to analyze the evolution of the radio luminosity function (LF) over the 
range $z=0-1.5$. Our sample of 346 radio sources represents
an increase of nearly a factor of five over H00 and enables 
us to provide a much more accurate determination of the $z=1$ radio 
LF. We use optical spectral classifications 
and X-ray characteristics to separate AGN dominated sources
from those that may be star formation dominated and to estimate 
the number density of ULIGs. An analysis of the submillimeter
and far infrared properties of the sample will appear in 
Wang, Barger, and Cowie 2004 in preparation.
We assume $\Omega_M={1\over 3}$, $\Omega_\Lambda={2\over 3}$,
and $H_0=65$~km~s$^{-1}$~Mpc$^{-1}$.

\section{Sample}
\label{secsample}

\markcite{richards00}Richards (2000) presented a catalog of 1.4~GHz
sources detected in a Very Large Array (VLA) map centered on the HDF-N
that covers a $40'$ diameter region with an effective resolution of 
$1.8''$. The $5\sigma$ completeness limit for compact sources in
the central region of the map is 40~$\mu$Jy.
\markcite{fomalont04}Fomalont et al.\ (2004) presented a
similar catalog within a $17'$ radius of
the 1.4~GHz VLA map ($5\sigma$ completeness of 25~$\mu$Jy)
of the SSA13 field (R.A.$=13^h 23^m$, decl.$=42^\circ 38''$).
In both cases, the completeness limits 
for more extended sources are several times higher.
The absolute radio positions are known to
$0.1''-0.2''$ rms. Optical imaging data for the two fields 
can be found in \markcite{capak03}Capak et al.\ (2003) and
\markcite{barger01}Barger et al.\ (2001), respectively.
\markcite{alex03}Alexander et al.\ (2003) presented
2~Ms {\it Chandra} Deep Field-North (CDF-N) X-ray images that
cover about 460~arcmin$^{2}$ around the HDF-N. 
Near the aim point, the data reach
limiting fluxes of $\approx 1.5\times 10^{-17}$ ($0.5-2$~keV)
and $\approx 10^{-16}$ ergs~cm$^{-2}$~s$^{-1}$ ($2-8$~keV).
\markcite{mushotzky00}Mushotzky et al.\ (2000) and 
\markcite{barger01}Barger et al.\ (2001) presented
100~ks {\it Chandra} exposures of the SSA13 field with
limiting fluxes 20 times higher than the CDF-N.

For the HDF-N region, we use a $9.3'$ radius
area around the approximate {\it Chandra} image center
R.A.$=12^h~36^m~51.20^s$ and decl.$=62^\circ 13'~43.0''$ (J2000.0).
The $9.3'$ radius is the largest circle that can be drawn
within the {\it Chandra} area. 195 of the 372 radio sources
lie within this circle. For the SSA13 field, we use a 
rectangular area of 175 arcmin$^2$ that contains 221 sources. 
We have so far spectroscopically observed 154 of these 
sources, chosen without regard to their radio or optical properties; 
thus, the effective area is 120~arcmin$^2$. Together these two 
restricted samples constitute our radio sample with 346 sources. 
All lie well within the half power radius of the radio observations, 
so the two fields should have relatively uniform flux selections.

%
%
\begin{inlinefigure}
\psfig{figure=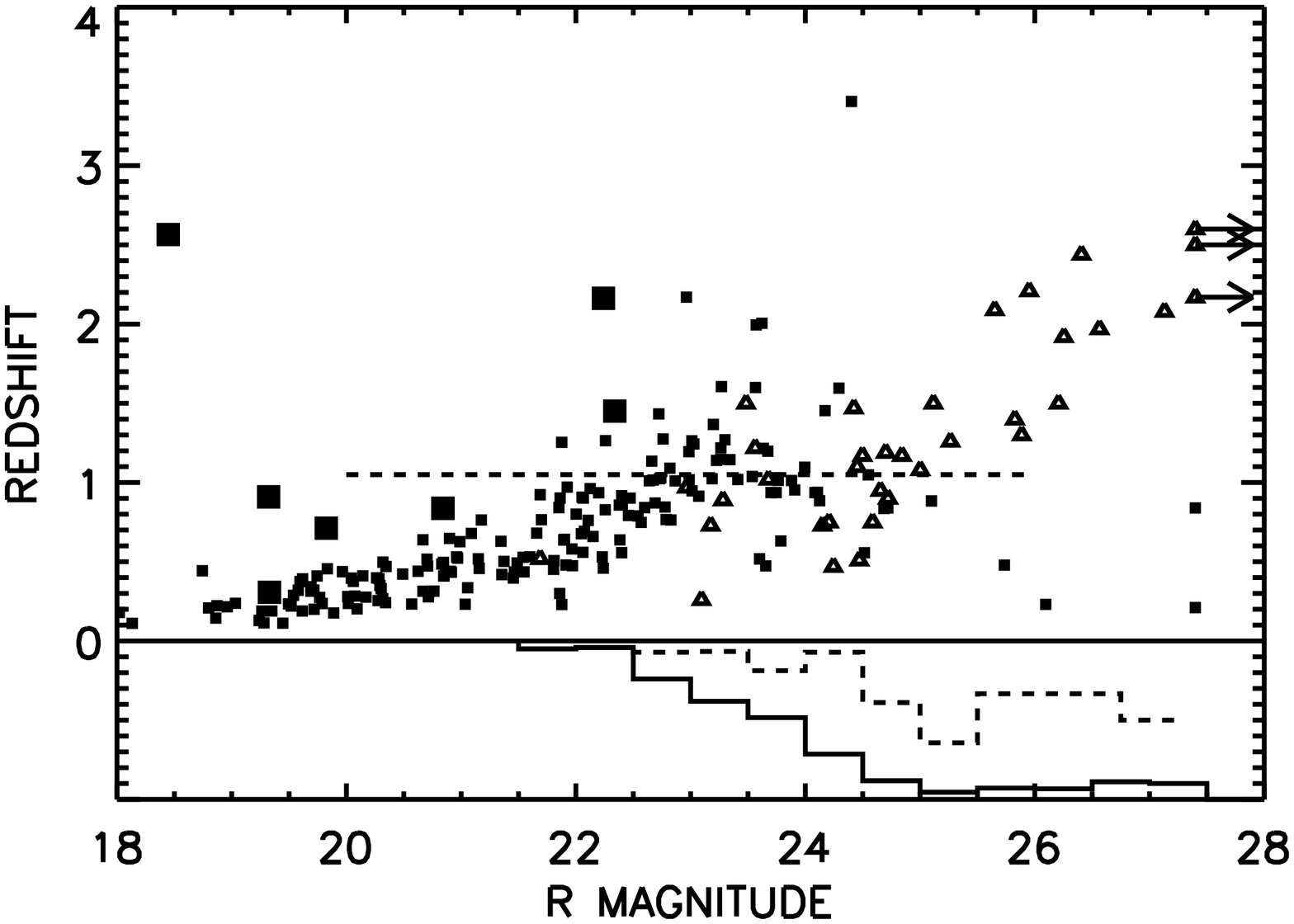,angle=90,width=3.5in}
\vspace{6pt}
\figurenum{1}
\caption{
Redshift versus $R$ magnitude for our radio sample,
excluding the two spectroscopically identified stars
({\it solid squares}---spectroscopic redshifts;
{\it open triangles}---photometric redshifts;
{\it large solid squares}---broad-line AGNs).
Dashed horizontal line shows the redshift ($z=1.05$) for which
a source with the median radio flux of 80~$\mu$Jy would have
a radio power that would correspond locally to a ULIG.
Solid histogram shows the spectroscopic completeness:
at $R<23$, $96\%$ (152 of the 162 sources)
are identified; at $R=23.5-24$, just over $50\%$;
at fainter magnitudes, only a handful with strong
emission lines. Dashed histogram shows the
completeness of the combined photometric and spectroscopic redshifts
in the HDF-N: 92\% of the $R<25$ sources have either a spectroscopic
or a photometric redshift.
\label{fig1}
}
\addtolength{\baselineskip}{10pt}
\end{inlinefigure}

We observed our radio sample with the
Deep Extragalactic Imaging Multi-Object Spectrograph
(DEIMOS; \markcite{faber03}Faber et al.\ 2003)
on the Keck~II 10~m telescope  and
the Low-Resolution Imaging Spectrograph
(LRIS; \markcite{oke95}Oke et al.\ 1995) on Keck~I.
We spectroscopically identified 192 of the 346 sources (55\%),
two of which are stars. Figure~\ref{fig1} shows spectroscopic redshift
versus $R$ magnitude ({\it solid symbols}) for our radio sample, as
well as the completeness of our spectroscopic identifications
versus $R$ magnitude ({\it solid histogram}). A source
with the median radio flux of 80~$\mu$Jy has a radio power corresponding
to a local ULIG at $z=1.05$ ({\it dashed horizontal line});
this corresponds to $R$ magnitudes in the range 22.5 to 25
(see Fig.~\ref{fig1}). Spectroscopic identifications
can be obtained for the majority of sources to $R=24$
and for some in the $R=24$ to $R=25$ range (see
solid histogram).

We can also estimate redshifts from photometric colors. For the 
HDF-N, we use the photometric redshifts of
\markcite{capak04}Capak et al.\ (2004), which are
based on ultradeep imaging in seven colors from $U$ to $HK'$
(\markcite{capak03}Capak et al.\ 2003) and were computed using
the Bayesian code of \markcite{benitez00}Ben\'{\i}tez (2000).
This code assigns odds that the redshift lies close to the estimated
value; we only use cases with odds above 90\%. 
The imaging data saturate at bright magnitudes, so we restrict 
to sources with $R>20$, leaving 161 photometric redshifts.
The photometric redshifts ({\it open triangles}) for sources
that do not also have spectroscopic redshifts are shown in
Figure~\ref{fig1}, as is the completeness of our HDF-N combined 
photometric and spectroscopic redshift sample
versus $R$ magnitude ({\it dashed histogram}).  
Eighty-three sources with $R>20$ have both spectroscopic
and photometric redshifts, and 92\% of these agree to within
a multiplicative factor of 1.25. This agreement is independent
of spectral type, except for the very small number of 
broad-line AGNs, where the failure rate is higher. The success rate 
of the photometric redshifts is consistent with that expected 
from the assigned odds. Since we do not have photometric
redshifts for the SSA13 sources, and since photometric redshifts do
not contain spectral type information, we only use the photometric 
redshifts to test the effects of completeness on the total LF.

For the spectroscopically identified sources, we classified 
the optical spectra into four spectral types,
roughly following the procedure used by
\markcite{sadler02}Sadler et al.\ (2002, hereafter S02)
to analyze low redshift 1.4~GHz samples. We classified sources
without strong emission lines (EW([OII])$<3$~\AA\ or EW(H$\alpha+$NII)
$<10$~\AA) as absorbers; sources with strong Balmer
lines and no broad or high-ionization lines as star formers;
sources with [NeV] or CIV lines or strong [OIII]
(EW([OIII]~5007~\AA$)>3$~EW(H$\beta)$) as Seyferts; and, finally,
sources with FWHM$>1000$~km~s$^{-1}$ as
broad-line AGNs. The breakdown by spectral type
in each field, and also by radio flux, is given in Table~1. 
S02 argued that the absorbers must be AGN dominated since 
star formation cannot produce the synchrotron emission in these galaxies.
They combined all but the star formers into a single AGN class.

The deep CDF-N X-ray data provide a check and a refinement of the 
classifications. We computed 
the $2-8$~keV rest-frame luminosity, $L_x$, for each source with a
spectroscopic redshift. In Figure~\ref{fig2} we
plot $L_x$ versus radio power for the
AGNs (classes 3 and 4; {\it triangles}), the star formers ({\it diamonds}), 
and the absorbers ({\it squares}). 
The sources classed as AGNs are  generally
X-ray luminous. Of the 40 AGNs, 29 are detected in the $2-8$~keV 
band, 26 of which have $L_x>10^{42}$~ergs~s$^{-1}$.
Such luminosities would securely identify them as AGNs
(\markcite{zezas98}Zezas, Georgantopoulos, \& Ward 1998;
\markcite{moran99}Moran, Lehnert, \& Helfand 1999). 
The remaining upper bounds are consistent
with all of the sources having $L_x>10^{41}$~ergs~s$^{-1}$. 

%
%
\begin{inlinefigure}
\psfig{figure=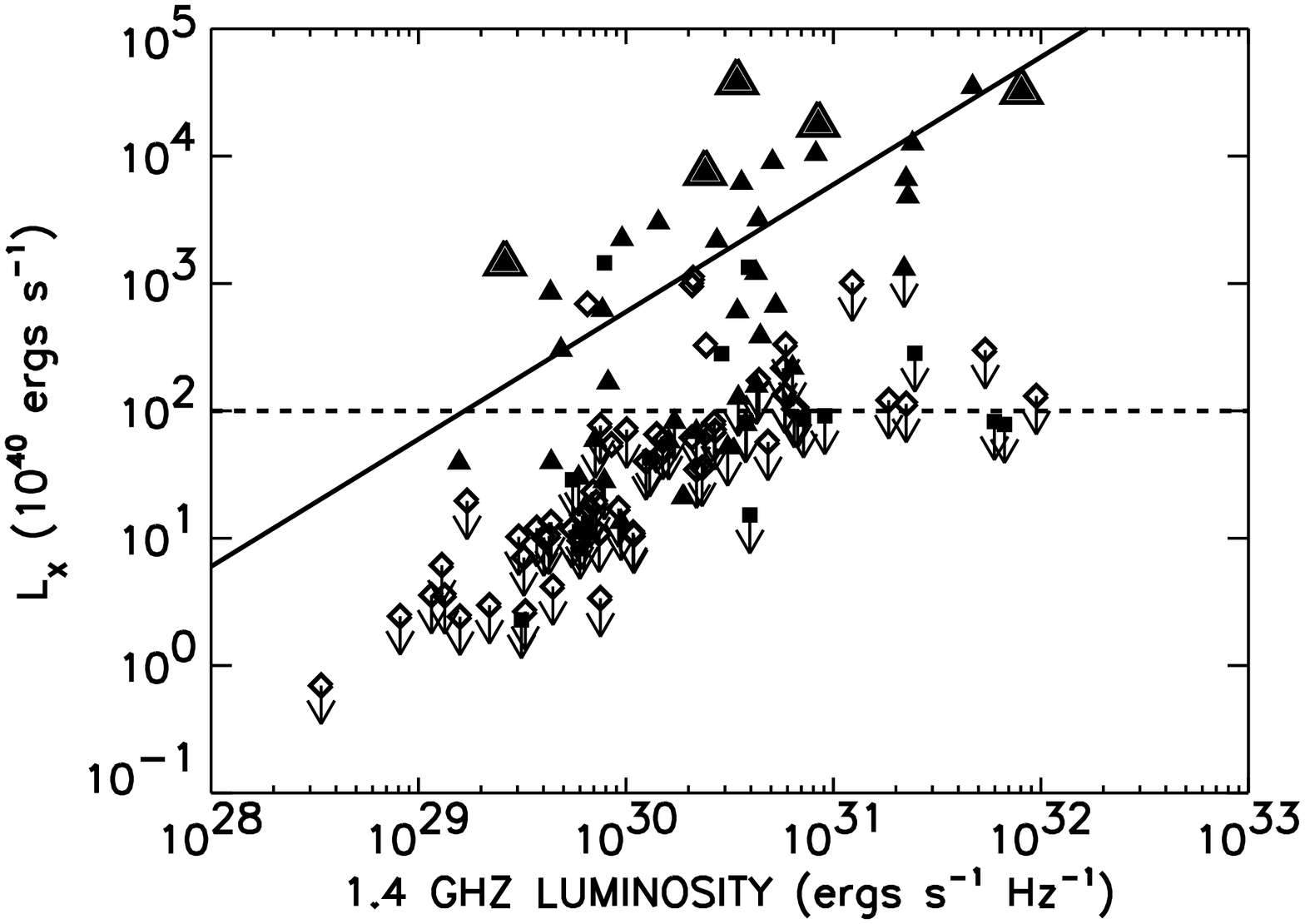,angle=90,width=3.5in}
\vspace{6pt}
\figurenum{2}
\caption{
$2-8$~keV luminosity versus 1.4~GHz power for the HDF-N radio sample.
Triangles show sources with AGN optical spectra
({\it large symbols}---broad-line AGNs; {\it small symbols}---Seyferts);
diamonds show star formers; squares show absorbers.
Arrows denote upper limits for sources not detected in the 2~Ms CDF-N.
Solid diagonal line is an approximate linear fit to the AGN population
and is just shown for guidance. Dashed horizontal line shows the
luminosity that would securely identify a source as an AGN.
\label{fig2}
}
\addtolength{\baselineskip}{10pt}
\end{inlinefigure}

By contrast, nearly all of the star formers have low X-ray 
luminosities. Only six of the 55 sources are detected in
the $2-8$~keV band, and only  
four have $L_x>10^{42}$~ergs~s$^{-1}$. The absorbers are 
the most unusual class. If they are AGN dominated, then they
are very underluminous in their rest-frame $2-8$~keV
luminosities relative to their radio power. This suggests
that they could be Compton-thick or near Compton-thick
sources. It may therefore be useful to maintain the distinction 
between the AGNs with clear optical signatures and the 
absorbers. However, to be consistent with S02,
we combine the two classes. The absorbers class has about half as many 
sources as the spectroscopic AGN class and a similar distribution 
of radio power. Thus, the AGN radio LF may be 
roughly corrected to a spectroscopic AGN LF 
by multiplying by $0.66$. Some sources that have
star-forming optical signatures may also correspond to this
type of source: X-ray quiet, radio detected AGN with no spectroscopic
AGN signatures. There may be ongoing star formation in the galaxy, but
the AGN dominates the radio power. Thus, the optically classified
star formers may only be an upper limit to the population
where star formation dominates the radio power. 

For our final classification, we classed all absorbers, star formers 
known to have measured (not limits) $L_x>10^{42}$~ergs~s$^{-1}$, 
and sources with AGN optical spectra as AGNs. We classed the remaining 
sources as star formers. This enables straightforward comparisons 
with the local results of \markcite{condon89}Condon (1989), 
\markcite{condon02}Condon, Cotton, \& Broderick (2002), 
\markcite{mac01}Machalski \& Godlowski (2001, hereafter MG01),
and S02.

\section{Discussion}
\label{secop}

We constructed the LFs for the 1.4~GHz power over the 
range $z=0.5-1.5$ using the $1/V_a$ method 
(e.g., \markcite{felten77}Felten 1977). This is shown for the 
total spectroscopic sample (AGNs and star formers;
{\it triangles}) in Figure~\ref{fig3}a, where it is compared with 
the local measurements of \markcite{condon89}Condon (1989; 
{\it solid line}). To investigate the effects 
of incompleteness, we computed the maximum total LF 
({\it squares}) for just the HDF-N using the much more complete 
combined photometric and spectroscopic samples and placing
all sources without a photometric or a spectroscopic
redshift into the $z=0.5-1.5$ bin. 
Many of these probably lie at higher redshifts, 
so this is an upper bound. Finally, we assumed that the
HDF-N sample was only 50\% complete in the $40-80$~$\mu$Jy
range, which approximates the completeness corrections
computed by \markcite{richards00}Richards (2000). The maximum 
correction for incompleteness is a factor of 2.6 at the lower 
radio powers but is relatively small at the higher radio powers.

In order to parameterize the evolution, we assumed
that the local portion of the LF allocated to star formers
in \markcite{condon89}Condon (1989) undergoes pure luminosity evolution
and that allocated to AGNs undergoes pure number density
evolution. We used these assumptions to calculate the expected LF 
in the $z=0.5-1.5$ interval. (The use of the S02 or the MG01 measurements,
which were constructed using more similar
methodologies to ours for classifying the galaxies, does not change 
the diagrams significantly.) This simple model provides a reasonable
fit to our measured LF, with the AGN number
density evolving as $(1+z)^{3}$, and the star-forming luminosities
evolving as either $(1+z)^{3}$ ({\it dotted line}), if we match the 
observed LF, or as $(1+z)^{3.8}$ ({\it dashed line}), if we match 
the maximum total LF. This systematic uncertainty is much larger than 
the statistical errors and primarily affects the evolution of the
star-forming LF because the uncertainties are larger at the lower
luminosities where this term dominates.
Our results for the evolution of the star-forming LF are
consistent with those determined from local samples by MG01
and \markcite{condon02}Condon et al.\ (2002),
who found a similar evolution of $(1+z)^{3\pm1}$. 
Our results are also broadly consistent with those of H00, who 
found a best pure luminosity evolution model with $(1+z)^{2.9}$.

In Figures~\ref{fig3}b and \ref{fig3}c we show the measured LFs 
in two narrower intervals
($z=0.6-0.9$ and $z=0.9-1.4$, respectively, chosen to 
match H00) and by spectroscopic class ({\it diamonds}---star formers;
{\it squares}---AGNs). For 
the star formers, we also show the results of H00 ({\it triangles}),
which, within the uncertainties, agree with the present measurements.
Our results are again broadly consistent with
the evolution described above (see Fig.~\ref{fig3}b caption for
description), although the AGN LF is slightly
high in the $z=0.6-0.9$ interval. This suggests that the 
power law evolution model may be too simple and that the
evolution of the AGN LF is faster at low redshifts and slower at high
redshifts. The star formers begin to show a more extended high 
luminosity tail in the $z=0.9-1.4$ interval.

%
%
\begin{inlinefigure}
\psfig{figure=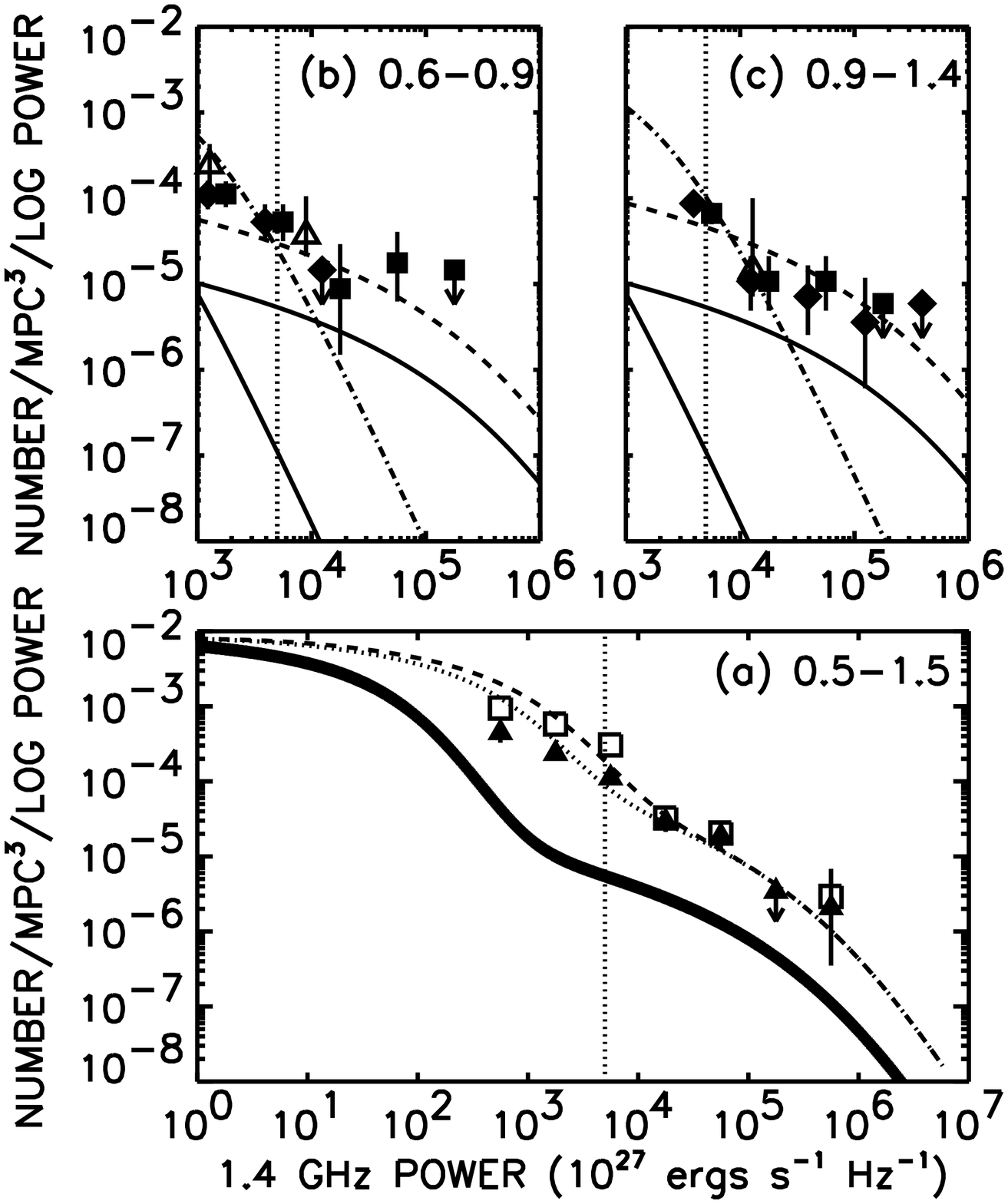,width=3.5in}
\vspace{6pt}
\figurenum{3}
\caption{
(a) Radio LF at $z=0.5-1.5$ for all the spectroscopically
identified radio sources ({\it triangles}), compared with
the local determination from Condon (1989; {\it solid line}).
Log denotes the base 10 logarithm. The $1\sigma$ uncertainties are
Poissonian. Squares denote the maximum total LF.
Vertical dotted line shows the equivalent
radio power of an $L_{FIR}=10^{12}~L_\odot$ source.
Dotted curve shows the expected LF if the local star-forming LF
undergoes pure luminosity evolution ($(1+z)^3$) and the local AGN
LF undergoes pure number density evolution ($(1+z)^3$).
Uncertainties are primarily systematic, and luminosities
evolving as $(1+z)^{3.8}$ provide a better fit
({\it dashed line}) to the maximum total LF.
(b) Radio LF at $z=0.6-0.9$ for the star formers ({\it diamonds})
and the AGNs ({\it squares}), to be compared with the local
star former ({\it steep solid curve}) and AGN
({\it shallow solid curve}) determinations from Condon (1989).
Triangles show the measurements of H00 (with a small correction
to make the geometries consistent) for the star formers.
Dot-dashed curve shows the expected LF if the local star-forming
LF undergoes pure luminosity evolution ($(1+z)^{3.8}$).
Dashed curve shows the expected LF if the local AGN LF
undergoes pure number density evolution ($(1+z)^{3}$).
(c) As for (b), but for the $z=0.9-1.4$ redshift interval.
\label{fig3}
}
\addtolength{\baselineskip}{10pt}
\end{inlinefigure}

The rapid evolution of the radio LF implies
that the number density of very luminous sources rises
rapidly in the $z=0-1$ interval for both the
star former and AGN classes though it may
be begining to slow at redshifts $z>1$.
In Figure~4 we show the number 
density of star formers with ULIG radio power ({\it solid squares})
in three redshift intervals.
The value from the  local star-forming radio LF is given by the 
open square. The evolution is reasonably well described by a $(1+z)^7$
evolution ({\it dashed line}). This evolution is similar
to that inferred by \markcite{bcr00}Barger et al.\ (2000)
from a comparison of the local number density 
of ULIGs and near-ULIGs to the $z=1-3$ number density of $>6$~mJy 
radio/submillimeter sources ({\it solid circle}).
If most of the submillimeter sources are dominated by star 
formation (\markcite{bcr00}Barger et al.\ 2000;
\markcite{chapman03}Chapman et al.\ 2003), the
match to the radio evolution is reasonable, if the radio
galaxies classified as star formers are indeed dominated by
star formation and the FIR-radio correlation holds to
these redshifts as suggested by recent results (Garrett 2002).

%
%
\begin{inlinefigure}
\psfig{figure=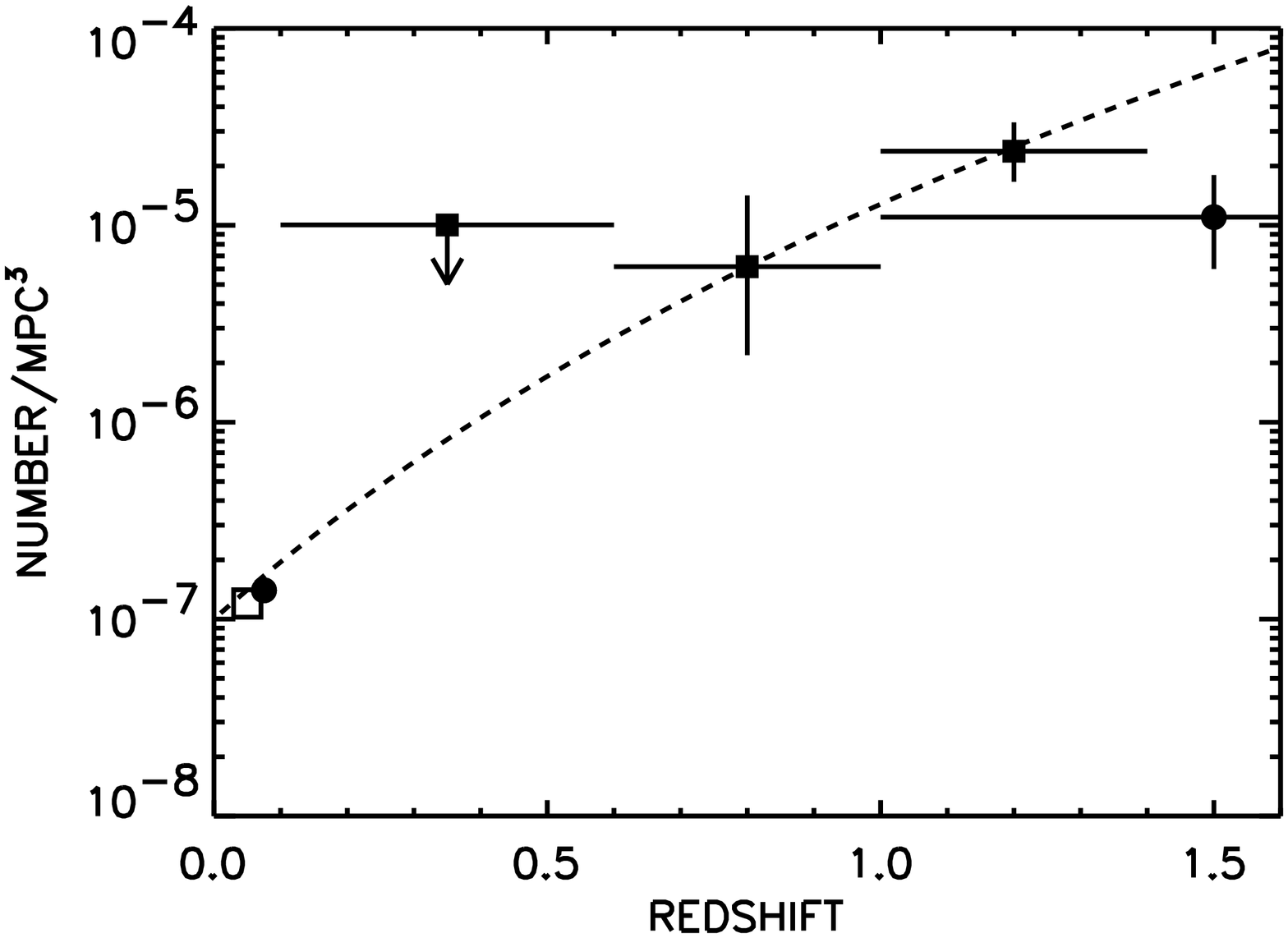,angle=90,width=3.5in}
\vspace{6pt}
\figurenum{4}
\caption{
Number density of star formers with ULIG radio power
in the redshift intervals $z=0.1-0.6$, $z=0.6-1.0$, and $z=1.0-1.4$
({\it solid squares}). Open square denotes the value from the
local star-forming radio LF.
Solid circles denote the local number density of ULIGs and near-ULIGs
and the $z=1-3$ number density of $>6$~mJy radio/submillimeter
sources (from Barger et al.\ 2000).
Dashed line shows a $(1+z)^7$ evolution.
\label{fig4}
}
\addtolength{\baselineskip}{10pt}
\end{inlinefigure}

\section{Summary}
\label{secsummary}

We have given a precise determination of
the high power end of the radio LF at high redshifts
($z\sim 1$). The LF was shown to evolve rapidly with 
redshift, both for galaxies with AGN spectra and for those 
with only star-forming signatures. This result is consistent 
with model expectations based on the local LF and the radio 
number counts (e.g., \markcite{condon89}Condon 1989).
The number of sources with radio luminosities that would 
correspond to ULIGs (based on local normalizations) matches 
the observed evolution from the local ULIG population to the 
radio/submillimeter sources at $z>1$.

\acknowledgements
We thank the referee for helpful comments that improved the manuscript.
L.~L.~C. gratefully acknowledges support from NASA grants GO2-3187B
and HST-GO-09425.03-A and NSF grant AST-0084816, and
A.~J.~B. from NASA grant HST-GO-09425.30-A,
NSF grant AST-0084847, NSF CAREER award AST-0239425,
the UW Research Committee, the Alfred P. Sloan 
Foundation, and the David and Lucile Packard Foundation.

\begin{deluxetable}{rrrrrrrr}
\renewcommand\baselinestretch{1.0}
\tablenum{1}
\tablewidth{0pt}
\tablecaption{Breakdown of spectral type by field and radio flux ($\mu$Jy)}
\small
\tablehead{CATEGORY & HDF & SSA13 & ALL & 25-50 & 50-100 & 100-200 & $>$~200 
}
\startdata
total  &    195  &  151  &  346 & 37 & 186 & 77 & 46 \cr
identified  &    115  &  82  &  197 & 20 & 98 & 51 & 28 \cr
stars  &    0  &  2  &  2 & 0 & 2 & 0 & 0 \cr
absorbers  &    15  &  14  &  29 & 1 & 16 & 4 & 8 \cr
star formers  &    57  &  49  &  106 & 10 & 51 & 33 & 12 \cr
Seyferts  &    36  &  13  &  49 & 8  & 23 & 12 & 6 \cr
broad-line  &    5  &  3  &  8 & 0  & 5 & 1 & 2 \cr
\enddata
\label{tab1}
\end{deluxetable}

\end{document}